\documentclass[letterpaper]{mn2e}

\def\simpropto{\lower.2ex\hbox{$\; \buildrel \propto \over \sim \;$}}
\def\ltsim{\lower.5ex\hbox{$\; \buildrel < \over \sim \;$}}
\def\gtsim{\lower.5ex\hbox{$\; \buildrel > \over \sim \;$}}

\usepackage{multirow}
\usepackage{rotating}
\usepackage{colortbl}
\usepackage{color}
\usepackage[fleqn]{amsmath}
\usepackage{amssymb}
\usepackage{amsfonts}
\usepackage{verbatim}
\usepackage{scalefnt}
\usepackage[percent]{overpic}

\def\gtorder{\mathrel{\raise.3ex\hbox{$>$}\mkern-14mu
     \lower0.6ex\hbox{$\sim$}}}
\def\ltorder{\mathrel{\raise.3ex\hbox{$<$}\mkern-14mu
     \lower0.6ex\hbox{$\sim$}}}

\newcommand{\apj}{ApJ}% Astrophysical Journal
\newcommand{\aap}{A\&A}% Astronomy and Astrophysics
\newcommand{\mnras}{MNRAS}% Monthly Notices of the RAS
\newcommand{\beq}{\begin{equation}}
\newcommand{\eeq}{\end{equation}}
\newcommand{\ba}{\begin{eqnarray}}
\newcommand{\ea}{\end{eqnarray}}
\def\spose#1{\hbox to 0pt{#1\hss}}

\newcommand{\comments}[1]{} %usage: \comments{}
 \title[Ultra-diffuse galaxies  without dark matter]{Ultra-diffuse galaxies  without dark matter}
 
\author[Silk]{Joseph Silk$^{1,2,3}$\\
$^{1}$ Institut d'Astrophysique de Paris, Sorbonne Universit\'es, UPMC Univ. Paris 06 et CNRS, UMR 7095, F-75014, Paris, France\\
$^{2}$ Department of  Physics \& Astronomy, The Johns Hopkins University, Baltimore, MD 21218, USA\\
$^{3}$ Beecroft Institute of Particle Astrophysics and Cosmology, Department of Physics,
University of Oxford,  Oxford OX1 3RH, UK
}

\begin{document}

\maketitle

\begin{abstract}
 I  develop a high velocity galaxy collision model to explain a rare but puzzling phenomenon, namely the apparent existence of {  ultra-diffuse galaxies} with little dark matter. Predictions include  simultaneous triggering of  overpressured dense clouds to form  luminous old globular clusters,  a  protogroup environment to generate high relative velocities of the  initially gas-rich  galaxies in the early universe, and 
 spatially separated dark halos, possibly detectable via gravitational lensing and containing relic low metallicity stars with enhanced $\alpha/Fe$ at ultralow surface brightness. 
 \end{abstract}


\section{Introduction\label{sec-intro}}

%\smallskip
%%textcolor{}
{  Ultra-diffuse galaxies (UDGs) } may have been discovered that have essentially no dark matter, at least in excess of their stellar content:
one  has been extensively studied
 \cite{2018Natur.555..629V}, and a second has recently been reported. Both are in the outskirts of the NGC 1052 group   \cite{2019arXiv190105973V}.  
 
 The existence of such dark-matter deficient UDGs is disputed on the grounds of uncertain  kinematic tracers \cite{2019MNRAS.484..245L} and distance estimators 
 %%%%%%%%%%% CHECK!!!!!!!{2019MNRAS.484..510N
% 
 %\cite{2018arXiv180610141T} and  \cite{2019MNRAS.484..510N},
\cite{2018arXiv180610141T,2019MNRAS.484..510N},
although these  arguments seem not to be definitive either for kinematics \cite{2019arXiv190103711D}
or for distance \cite{2019arXiv190202807V}. {  The distance issue provides the  outstanding uncertainty in assigning dark matter mass, but should be resolvable with new HST data.}

Suppose such DM-deficient UDGs exist. I argue that 
this is a challenge for galaxy formation theory. Our most likely candidates, dwarfs formed via  gas condensation in tidal tails, TDGs,  are known to be deficient in dark matter. However  even aged counterparts of typical tidal dwarfs would be  more gas-rich  and significantly brighter than the UDGs \cite{2015A&A...584A.113L}. 

Generation of the extremely low surface brightnesses observed  requires  either the  formation of stars in tidal dwarfs, observed to be efficient \cite{2019arXiv190310789F}, followed by strong tidal heating, which seems a  contrived sequence of events, or a very low efficiency star formation rate in the first place. Moreover, the associated bright globular clusters are an important clue that is suggestive of a somewhat more exotic formation pathway \cite{2018ApJ...856L..30V}, as  proposed in the mechanism  that I describe below.

There is also a  kinematic argument against a tidal interpretation of DM-deficient UDGs.
Tidal heating of compact dwarfs indeed offers a means of  removing dark matter outside a scale length \cite{2010MNRAS.406.1290P}. If the dwarf is cored, this can even result in formation of a  UDG \cite{2018MNRAS.480L.106O}.
In this case, the SFE would be  normal but the  observed low surface brightness is due to tidal heating.

One consequence would be predominantly radial stellar orbits. 
However, any gradient, and in particular the ensuing relaxation to the observed cold  and apparently relaxed kinematics,  is claimed  to not  be supported by the dynamical data and  modeling of orbital parameter space in the UDGs \cite{2018ApJ...863L..15W}. 
%%%%%%%%%%
{ Moreover, low surface brightness features are ubiquitous in the NGC 1052 group and do not support the case for tidal interactions of the two UDGs with their possible hosts
\cite{2019A&A...624L...6M}. In fact the dynamical data is far too sparse  in terms of numbers of data points 
to reach any definitive conclusion about the role of tidal interactions.}

{ The allegedly dark matter-deficient UDGs are relatively luminous for low surface brightness dwarfs, and one might also consider initial conditions as a formation pathway, due to high initial spin \cite{2016MNRAS.459L..51A}.
Their ultra-diffuse nature may also be due to tidal heating \cite{2019MNRAS.485..382C} or to early supernova-driven gas outflows  \cite{2019MNRAS.486.2535D}. However these 
mechanisms do not obviously account for the dark matter deficiency, nor most significantly for the bright globular clusters.}

 The existence of dark matter-deficient UDGs  is a rare phenomenon, and I argue here that it is due to a mini-Bullet cluster-like event, involving the high velocity collisions of gas-rich dwarfs occurring  at early  epochs in a protogroup or protocluster environment.  Gas dissipation and low efficiency star formation, along with DM separation
  from the baryonic gas component  and consequent expansion of the residual stellar systems,   are a natural consequence of high velocity infall and  collisions between gas-rich dwarfs.
  
 Ram pressure stripping, generally more significant than tidal stripping in the group environment \cite{2018arXiv181110607J}, provides a means of depleting   the low density gas reservoir whose accretion ordinarily enhances star formation in isolated galaxies.
 % Such high velocity collisions occur naturally in the formation phase of galaxy groups and clusters. The rarity of the phenomenon is attributed to the gas-rich content of the colliding dwarfs, which requires an early epoch. and high  infall velocities.  High velocity collisions as well as ram pressure naturally lead to low star formation efficiency. 
At the same time, dense self-gravitating clouds within the colliding systems are highly overpressured and collapse to form  protoglobular-like star clusters  at relatively high star formation efficiency. I give simple arguments below to support these conjectures.
 %One obvious hurdle is that the  onset of significant star formation during infall to the protogroup/protocluster environment must be delayed.  

\section{High velocity collisions of gas-rich dwarf galaxies}

Collisions and mergers are common between dwarf galaxies even at high redshift.
The Fornax dwarf is a classic example of past mergers as evidenced in shells 
\cite{2017MNRAS.465.3708D}
and complex substructures 
\cite{2018arXiv180907801W}.
The ultra-diffuse dwarfs require special initial conditions. I suggest here that they, and especially the supposedly dark matter-free UDGs, are the consequence of high velocity gas-rich collisions in the past, typically in proto-group 
%or proto-cluster 
environments. In such conditions, the induced star formation is highly inefficient. Hence if the colliding dwarfs had not previously undergone significant star formation, only UDG-like systems would be produced. Moreover at  the highest collision  velocities  envisaged ($v_{coll} \sim 300$ km/s),  
%typical of proto-clusters, 
the 
non-dissipative dark matter would be well separated from the dissipative gas components that  eventually merge and form stars, and the resulting  decrease in self-gravity further contributes to expansion of the newly formed stellar system. { Nor is the present mechanism limited only to dwarfs: high velocity collisions of more massive gas-rich systems are equally capable of forming UDGs.}

\subsection{Star formation\label{sec-SF}}

Star formation is generally inefficient, at the 1\% level in GMCs and in nearby star-forming galaxies. This is directly measured  for galaxies via the Schmidt-Kennicutt relation.  
However the observed diversity in the star formation efficiency (SFE)  of  individual GMCs ranges from of order  50\% in dense  star-forming cloud  cores to 0.01\% in the most quiescent giant molecular clouds \cite{2019arXiv190200518G}. Understanding this  variance is where all the physics resides.

A promising approach to understanding the diversity of SFEs  models supersonic turbulence in the presence of magnetic fields.  This  exponentially reduces the SFE \cite{2012ApJ...759L..27P}.
The SFE per cloud free-fall time is likely to be very low because the high velocity collision/merger generates turbulence  and shear in the gas. Typical cloud collisions are oblique and induce shear. This is very effective at suppressing fragmentation  in the diffuse gas \cite{2018MNRAS.474.1277A}. The local dynamical time will be much less than the local free-fall time  in colliding clouds. 

The SFE has been shown to be exponentially suppressed in such situations, provided that interstellar  magnetic fields are initially present to cushion the impact. The fields are amplified  by stretching and compression according to MHD  simulations \cite{2012ApJ...759L..27P}. The SFE is found to be reduced by a factor $\rm exp(-t_{ff}/t_{dyn}), $ where $t_{ff}=\sqrt {3\pi/32G\rho_{cl}},$  $t_{dyn} =L/2\sigma_{turb},$  $\rho_{cl}$ is the mean density, $L$ is the cloud size, and $\sigma_{turb}$ is the 3-d turbulent velocity dispersion in the cloud.  
%One can thereby account for the low surface brightness of UDGs. 

\subsection{Application to UDGs}

There are four key  issues to be explained in UDGs.  I focus on the two examples which  allegedly 
lack dark matter. 

A. Why do some UDGs have no DM? 

B. Why  is the SFE so low? 

C. Why do they have large  core radii and low surface brightnesses? 

D. And why are there many  bright globular star clusters?

The mini-Bullet cluster hypothesis potentially explains all four issues. 

Gas dissipates during a gas-rich merger  
%as. $ t_{cool} <   t_{dyn}$.   In fact, a stronger condition is found empirically via
in analogy with  cold cloud precipitation from the circumgalactic medium  into the galaxy.  Ensuing cooling and star formation  can occur once  the gas cooling time-scale is less than 10 times the local free-fall time \cite{2015ApJ...808L..30V}.
Hence the gas agglomerates and eventually is Jeans unstable. The Jeans mass increases by a factor $\mathcal{M}^2.$ 
%Hence the core is large. 
The high Mach number $\mathcal{M} $ and low gas density,   more specifically the large ratio $t_{ff}/t_{dyn},$ guarantees low SFE.  

Kelvin-Helmholtz instabilities are important during the gas-rich collision of merging galaxies. The ensuing turbulence leads to a large core radius for the distribution of triggered star formation. Disruption of clouds by Kelvin-Helmholtz instabilities is suppressed at high Mach number \cite{2015ApJ...805..158S}, hence star formation is not completely quenched. 

High column density clouds are not significantly accelerated and are compressed into dense, potentially star-forming filaments \cite{2016ApJ...822...31B}. 
Kelvin-Helmholtz mixing between the hot wind and cold cloud helps preserve cold gas clumps \cite{2018MNRAS.480L.111G}.
The role of self-gravity however has not been incorporated into these simulations.

For SFE suppression,  one needs  a combination of long 
$t_{ff } $ and short $t_{dyn}$.  Consider the  Central Molecular Zone    of our galaxy. 
%Low SFE  is observed in molecular clouds near the Galactic Centre, where turbulence is enhanced with Mach numbers of order 10 \cite{2019arXiv190310617D}.
Here  the  inferred turbulence Mach number  $\mathcal{M} \sim 10$   \cite{2019arXiv190310617D} and the SFE is low \cite{2019arXiv190107779L}.
%%%%%%

Comparison of  $t_{cool }$ with $t_{dyn}$ in high velocity gas cloud collisions,  eg with Mach numbers in the range  10-100, suggests that  only the densest clouds cool within a dynamical time.
High velocity, of order 100 km/s collisions, are needed for inefficient star formation in gas-rich dwarfs. This points to dwarf  formation in  protogroups or especially in protoclusters as providing high velocities  due to gas-rich galaxy  infall into  a gas-rich environment. 

If  the infall velocity  is too high,  cooling becomes too  inefficient, e.g.  at collision velocity $v_{coll}\sim 1000$ km/s for gas aggregation and star formation.  Group environments are the sweet spot, possibly also  in protoclusters.  The infalling galaxies need to be gas-rich, hence the protogroup/protocluster environment is optimal.

The combination of low SFE and expansion should account for the low surface brightnesses.
Firstly, a factor of 10 or more  comes from the reduced SFE. One can infer  this from the galaxy main sequence or the Schmidt-Kennicutt  relation.
 The    SFE is  expected to be  proportional to the  ratio of $t_{dyn}$  to $t_{ff}$, which is of order $ \sigma_{turb}/v_{coll}\ltsim 0.1$.
So this gives a total reduction of up to 3 magnitudes in surface brightness.

Secondly, the orbital momentum adiabatic  invariant implies that the outflow has a dramatic effect on the final distribution of newly formed stars. I assume that
$r\sigma_\ast$ is constant, where $r$ is the stellar orbit perihelion and $\sigma_\ast$ is the stellar velocity dispersion.
 Hence for a cloud expansion velocity $\sim 10 \sigma_*$,  one can reduce the  final surface brightness by up to $\sim 1000.$

 \subsection{Globular cluster formation}
It is a challenge to generate extremely low surface brightness dwarfs which nevertheless have  significant populations of (possibly anomalously bright) globular clusters. 
%I argue that  high velocity shocks and associated turbulence both lower the mean SFE and simultaneously overpressurise dense gas clumps to form compact star clusters. %Feedback from IMBH will also play a role in expanding the gaseous precursors of the stellar  cores. 
 High pressure environments are essential to globular cluster formation 
\cite{1997ApJ...480..235E}.
The extreme high pressure environment  $p_{coll}$ achieved in a fast collision should lead to enhanced masses of newly formed globular clusters at a given cloud density. 

For example, the mass of a self-gravitating cloud is
$M_{cloud}=(9/16)(\pi)^{-1/2}(p_{coll}/G)^{3/2}\rho_{cl}^{-2}.$ Cloud collision velocities in excess of 50-100 km/s are argued to account for the dynamical and optical characteristics of globular clusters \cite{1993ApJ...404..144K} and confirmed in more recent simulations \cite{2004ApJ...602..730B}.

 Moreover, the shear induced by encounters  imparts angular momentum, and another consequence is that  enhanced shear drives fission of dense molecular clouds. For example, 
 binary globular clusters constitute about 20\% of the mostly young globulars in the LMC \cite{2019JPhCS1127a2053P} and likely formed in high velocity cloud collisions \cite{1997AJ....113..249F}. They are destined to merge via torquing in  the local tidal field \cite{2016MNRAS.457.1339P} to further enhance the population of massive globular clusters
 and potentially account for   the puzzling age gap in LMC globulars
 \cite{2004ApJ...610L..93B}.  Collision-enhanced pathways boost the numbers of anomalously  bright globular clusters as possibly observed in  the two DM-free UDGs. 
Numerical simulations of colliding gas streams are found to provide a plausible  
 formation mechanism for dark matter-free  globular star clusters, in a study that appeared after this paper was initially submitted  \cite{2019arXiv190508951M}.
%%%%%%
%%%%%%
%%%%%%
Observations are equally suggestive. As  many as 60 extraplanar star-forming  HII regions with masses $10^3-10^5 \rm M_\odot$
have been mapped outside the optical disk in the gas tidally stripped from a single disk galaxy,  likely progenitors of  globular clusters or ultra-compact   dwarf galaxies \cite{2018A&A...615A.114B}. 

Ram pressure stripping indeed leads to induced star formation as  in the jelly fish phenomenon \cite{2018ApJ...866L..25V}.
 The effects of a high velocity collision should  enhance such effects, reducing the SFE in diffuse gas but enhancing  self-gravity  and hence fragmentation in dense clouds
 
 Dense clouds are not significantly accelerated by ram pressure.  There is a trade-off  in their eventual fate  as Kelvin-Helmholtz instabilities on   cloud boundaries lead to mixing and heating \cite{1990MNRAS.244P..26B} with consequent stabilization against gravitational fragmentation. However 
 inclusion of self-gravity into the simulations would argue strongly for compression and triggering of star formation, as inferred from simulations  of jets \cite{2012MNRAS.425..438G} and winds \cite{2017ApJ...844...37D}.  Thermal conduction most likely is not important in suppressing the instabilities once magnetic fields are included \cite{2018MNRAS.473.5407M}.

\subsection{Frequency \label{sec-Frequ}}

The dwarf galaxy merger rates are low at the present epoch, of order 0.03/Gyr according to recent simulations \cite{2015MNRAS.449...49R}.
Some  occur as evidenced by morphological distortions such as shells but these are relatively rare, although close dwarf pair frequency increases with reduced stellar mass
\cite{2018MNRAS.480.3376B}.
However redshift scaling suggests that they will be much more common at higher redshift, when galaxies are gas-rich. For example, over a wide range of masses,
 the major merger fraction increases as $\sim (1+z)^6$ to $z\sim 2$, peaking near the maximum in star formation rate density, and then flattens towards higher $z$ \cite{2017A&A...608A...9V}.
{  For dwarfs, some 10\% experience major mergers within the host group virial radius since $z
 \sim 1$ \cite{2014ApJ...794..115D}.
Faint shells are seen at several effective radii around dwarfs in a deep survey of the Virgo cluster, indicative of  high velocity interactions and recent equal mass gas-free mergers \cite{2017ApJ...834...66P}. This suggests that such interactions should have been  frequent at early epochs when assembly is dominant  in a gas-rich environment at first infall beyond the group or protocluster virial radius.}

To try to crudely quantify the expected collision velocity, I assume that the  probability distribution of finding  the most massive subhalos  with a velocity 
larger than $v_{sub}$ is fit at $z\sim 0.5$ by  
$$
log f(> v_{sub}) = -\left({v_{sub}/v_{200}\over{1.6}}\right)^{3.3},
$$
where $v_{200}$ is the cluster velocity dispersion \cite{2006MNRAS.370L..38H}.
This calculation, originally performed for the Bullet cluster and assuming gaussianity,  underestimates the high velocity tail, which is effectively nongaussian \cite{2015MNRAS.450..145B}. Rare collisions at 2-3 times the mean protogroup/cluster velocity should  suffice to maintain a low SFE while at the same time maintaining a high enough dissipation rate to separate the gas from the weakly interacting  DM concentration.
One  requires  protoclusters  and protogroups containg such infalling clouds to be sufficiently rare  in terms of gaussian theory in  order to have gas clouds infalling without previously making excessive numbers of stars. 
%%%%%
%%%%%
%%%%%

\subsection{The IMBH connection}
Quenching and gas outflows are likely to be driven by central  intermediate mass black hole (IMBH) AGN in dwarfs, although the details are still poorly known. Angular momentum transfer leads to extreme gas densities. IMBH formation is increased as fragmentation is reduced. Shear is  further increased in high velocity mergers, and this in turn drives Kelvin-Helmholtz instabilities. These effects can  also lead to suppression of fragmentation \cite{1997ApJ...483..262V}. 

 In fractal clouds, effects of  compression allow   retention of high density nuclei that survive and allow mass loading into the hot outflows
\cite{2019arXiv190106924B}.
However radiative cooling eventually results in fragmentation until the clouds are fully dissolved in the hot wind \cite{2019MNRAS.482.5401S}. Hence IMBH formation occurs without excessive fragmentation into stars.

IMBH may be ubiquitous in dwarfs, as suggested by theoretical considerations 
\cite{2017ApJ...839L..13S}
and by recent surveys especially using IR diagnostics
% \cite{2017A&A...602A..28M} and \cite{2018ApJ...858...38S}.
\cite{2017A&A...602A..28M,2018ApJ...858...38S}.
It is possible that IMBH form before significant fragmentation into stars.
If IMBH  indeed form early, they would help suppress Pop III star formation via generation of UV Werner band flux. 
Various alternative schemes have been suggested for suppressing fragmentation and early star formation in dwarfs, including 
 disk gravitational stability \cite{2014MNRAS.445.1549I} and magnetic  disk levitation \cite{2017MNRAS.464.2311B}.

\section{Discussion}

%\smallskip\

I have argued that high velocity shocks and associated turbulence both lower the mean SFE and simultaneously overpressurize dense gas clumps to form compact star clusters. 
%There are other implication of the dwarf collision model that merit  brief comments. I note that 
The collision model requires the stars and the globular clusters to have similar old ages, around  10 Gyr corresponding to the gas-rich phase of the newly formed dwarfs. 
Violent star formation is expected to induce $\alpha/Fe$-rich chemistry, because only SNII will play a role in enrichment.
There will not be time for SNIa superovae to contribute to stellar metallicities, hence one expects  both the UDGs and the associated globular clusters, formed contemporaneously in the present model,  to have 
$[\alpha/Fe]\sim 0.3$ and $[Fe] \ltsim -1$ in the  UDGs that lack DM.

Some $\sim 10\%$ of the stars are expected to be ejected in the GC mergers, according to the simulations, and 
UDG halos should contain some old  $\alpha-$rich metal-poor stars. One cannot resolve individual stars but the globular clusters in the UDGs can be assessed for age, metallicity and any $\alpha$-element excess by  population synthesis analysis of stacked spectra \cite{2018ApJ...856L..30V}.

%\subsection{Environment}

The frequency of tidal  tail galaxies  as well as  of  UDGs should be  enhanced in groups where galaxy mergers and  collisions are likely to be  prevalent.  
The gas-rich environment at early epochs suggests that  protogroups may be especially promising for forming UDGs because one has  a gas-rich environment, an enhanced frequency of dwarfs, and a high velocity tail in the galaxy velocity distribution. 
UDGs without  DM should therefore be  old and in dense environments if of collisional origin.

In summary, if indeed UDGs exist with little dark matter, there are at least two possible mechanisms for explaining their origin.
These are origin via tidal tail instability in the vicinity of a massive host galaxy or via high velocity collisions of dwarfs. A unique aspect of the latter model  is  that it provides a plausible way of   forming UDGs with associated globular clusters of enhanced brightness.  

The predicted presence of displaced dark halos
that should contain residual metal-poor stars
is potentially detectable by weak lensing. Another is the morphology of the globular cluster distribution  surrounding the two candidate  dark matter-deficient UDGs in the NGC 1052 group.  { Yet another relic would be the displaced gas that has mostly not formed stars. Presumably, this is sufficiently diffuse to remain hot and only be accessible by x-ray observations or absorption towards distant quasars.}

A third possibility would involve a supermassive black hole (SMBH), hosted by the central galaxy  and  fueled by gas accretion and generating a powerful jet  or outflow.  Star formation is plausibly triggered in overpressured dense halo clouds.
% \citep{2012MNRAS.425..438G,2017ApJ...844...37D}. 
 This is   observationally plausible, cf. local systems such as 
  %given the paucity of examples of jet triggering of formation of dwarf galaxies, 
 Minkowski's object \cite{2017ApJ...850..171F} and Cen A  \cite{2017A&A...608A..98S},  along with more remote examples \cite{2018NatAs...2..179C}. However, 
% For example,  there is  scope  for the erosive role of Kelvin-Helmholtz  and other instabilities. 
an improved understanding of  massive cloud survival is  needed to decide whether  jet entrainment  of ambient gas clouds is indeed  followed by  compressionally-triggered star formation and formation of UDGs, or rather by cloud erosion. 

 The presence of a SMBH in the host galaxy NGC 1052 \cite{2019MNRAS.tmp..693F}
 of the 
two likely DM-free UDGs may support  the possible role of  early jet-triggering. Morphology and kinematics of the UDGs,  and especially of their associated globular clusters, could help decide this issue.

%%%%%%%%%%%%%%%%%

%{ \color{}{
IMBH outflows have recently been detected  in dwarfs that contain AGN \cite{2019arXiv190509287M}.
The outflows are potential creators of UDGs. AGN outflows in the group environment can be effective at several virial radii in gas removal from dwarfs \cite{2018MNRAS.473.5698D}
but could also trigger star formation in dense gas clumps. Self-gravity has not hitherto been included in the simulations to properly address this point.

High velocity clouds  are another environment where star formation may have occurred inefficiently as in UDGs.
Recent detections of associated ultra-faint galaxies with comparable amounts of gas and old stars, but no evidence of recent star formation 
\cite{2019AJ....157..183J},
 suggest that  these systems that are infalling to the MWG or Local Group might be  relics of a high velocity encounter that induced star formation.. Only the residual gas core remains along with associated  stars that are diffusely  distributed.  A similar phenomenon has been found for  an ultrafaint galaxy detected in deep imaging of the Virgo cluster \cite{2018MNRAS.476.4565B}.  These are generally old stellar systems, and so it is likely that the observed gas is a  relic of a far more extensive gas component, favoring low star formation efficiency.

Primordial black holes  (PBHs) of stellar mass are a currently popular form predicted for a component of dark matter  via explanations of the LIGO event rates \cite{2017PhRvD..96l3523A}. 
Destruction of UDGs has been  suggested  \cite{2016ApJ...824L..31B} as an argument against stellar mass   PBHs  being a dominant component of the dark matter. 
However it is possible that 
PBHs, if sufficiently numerous,  could equally create UDGs via their dynamical effects on dwarf galaxies. 

With improved  treatment of self-gravity in numerical simulations,   survival  of UDGs is an option that should be considered. Simulations of  the  Fornax dwarf galaxy find that the observed  globular clusters are not necessarily incompatible with a NFW dark matter profile \cite{2019MNRAS.485.2546B},
in contrast to earlier discussions of the core/cusp problem for dark matter \cite{2017ARA&A..55..343B}. The action of dynamical friction leading to infall and merging  of globular clusters is sensitive to initial orbital parameters, which in turn come from cosmological simulations. 

More generally, the effects of tidal heating can be significant. They are argued to play an important role in  reducing the number of dwarfs predicted in the Local Group in the standard CDM model \cite{2017MNRAS.471.1709G}.
 It remains to be seen whether such effects could also  lead to survival of objects morphologically resembling UDGs.
%}}
%%%%%%%%%%%%%%%%%

\section*{Acknowledgements}

%\bigskip
%\begin{acknowledgments}
I thank Annette Ferguson, Sadegh Khochfar  and Maxime Trebitsch for discussions. 
%\end{acknowledgments}

\bibliographystyle{mn2e}
\bibliography{../../biblio}
%

% % % % % % % % % % % % % % % % % % % % % % % % % % % % % % % % % % % % % % % % % %

\end{document}